\documentstyle[11pt,epsf,newpasp,psfig,twoside]{article}
\reversemarginpar

\begin{document}
\title{Galactic Center:
       Implications of Recent {\it Chandra}
       Observations for Spherical Accretion Models of Sgr A*} 
\author{Robert F. Coker}
\affil{Department of Physics and Astronomy, University of Leeds, 
              Leeds LS2~9JT  UK}
\author{Sera Markoff\altaffilmark{1}}
\affil{Max-Planck-Institut f\"ur Radioastronomie, 
Auf dem H\"ugel 69, 53121 Bonn, Germany}
\altaffiltext{1}{Humboldt Research Fellow}

\begin{abstract}
At the center of the Milky Way lurks a unique compact nonthermal radio
source, Sgr A*.  It is thought to be powered by a $2.6\times10^6$
solar mass black hole that is accreting the stellar winds from the
numerous early-type stars that exist in the central parsec.  However,
until recent high resolution {\it Chandra} observations, Sgr A* had
never been unequivocably detected at wavelengths shorter than the
sub-millimeter.  We present a spherical accretion model which is
consistent with both the flux and steep spectral shape of the X-ray
emission from Sgr A*.
\end{abstract}

\section{The Spherical Accretion Model}

Sagittarius A* (Sgr A*), is located at the dynamical heart of the
Milky Way (Reid et al. 1999; Backer \& Sramek 1999).  Proper motion
studies of stars in the Galactic Center (GC) indicate that Sgr A* is a
$\sim 2.6\times 10^6 \;M_{\sun}$ black hole (e.g., Ghez et
al. 2000; Genzel et al. 2000).  There are more than two dozen known
early-type stars in the central parsec (Genzel et al. 1996); combined,
they produce winds that total $\sim 10^{-3}\;M_{\sun}$ yr$^{-1}$
(Najarro et al. 1997).  In the classical Bondi-Hoyle scenario (Bondi
\& Hoyle 1944), one would expect Sgr A* to be accreting a substantial
fraction of these winds ($\sim 10^{-4}\;M_{\sun}$ yr$^{-1}$).  This
picture is consistent with hydrodynamical simulations (Coker \& Melia
1997), although the precise accretion rate depends on the spatial
distribution of the wind sources.

In our spherical accretion model (Coker \& Melia 2000), the radio
spectrum of Sgr A* is due to magnetic bremsstrahlung.  However, as the
accreting winds fall down the gravitational well of Sgr A*, the gas
will also emit thermal bremsstrahlung, resulting in a significant
X-ray flux.  As pointed out by Quataert, Narayan, \& Reid (1999),
X-ray observations can then be used to place limits on the accretion
rate.  In this work, we modify our model to satisfy the spectral index
and flux limits of recent X-ray observations.

Initial ROSAT X-ray observations provided an upper limit on the
luminosity of Sgr A* of $\sim 4~L_{\sun}$ in the 0.8--2.5 keV range
(Predehl \& Zinnecker 1996).  More recent, higher resolution {\it
Chandra} observations show a source that is coincident with Sgr A* to
within 1\arcsec~(Baganoff et al. 2000).  The luminosity of the source
is $\sim 2 L_{\sun}$ in the 0.5--10 keV range, suggesting that the
ROSAT observations included either other sources or significant
diffuse emission.  Such a low luminosity leads to an upper limit on
the accretion rate of $\sim 10^{-5}\;M_{\sun}$ yr$^{-1}$ (Quataert et
al. 1999), significantly less than the predicted Bondi-Hoyle accretion
rate.  The X-ray spectrum of Sgr A* appears to be very soft, with a
photon index of $\sim 2.75^{+1.25}_{-1.00}$ (Baganoff et al. 2000),
while the spectrum of thermal bremsstrahlung emission is fairly flat
until it cuts off exponentially at $h\nu \sim kT$.  Since the radio
emission requires relativistic temperatures, it is unlikely that the
soft X-ray spectrum is due to thermal bremsstrahlung alone.

As discussed in Coker \& Melia (2000), the accretion flow that arises
from nearby stars is assumed to be quasi-spherical, with no disk
formation.  We solve for the temperature and velocity profile, taking
account of magnetic heating and radiative cooling.  We find a
sub-equipartition field at large radii with a nearly constant ratio of
magnetic to thermal pressure,
%an approximately thermal equipartition
%field at large radii, 
a constant field in a transition region, and a
rapidly increasing field close to the event horizon.  A dynamo at
small radii is required to reproduce the ``sub-mm bump'' seen in the
spectrum of Sgr A* (Serabyn et al. 1997; Falcke et al. 1998).  Such a
dynamo could easily be the result of instabilities in the flow which
convert thermal energy to magnetic energy as the gas begins to
circularize (Melia, Liu, \& Coker 2000).

This approach, however, is quite an oversimplification.  In reality,
even though the dynamo will transport angular momentum outwards
(Balbus \& Hawley 1991), the flow is likely to have sufficient
residual angular momentum to form a Keplerian flow at $r \la 50 r_s$
where $r_s \equiv 2 GM/c^2$ is the Schwarzschild radius.  However,
unless the net accreted angular momentum vector is constant in time
(see, e.g., Genzel et al. 2000), this flow is not likely to settle
into a true accretion disk.  We visualize an unhindered spherical flow
at large radii, a smooth transition region where the flow
circularizes, and an inner ($r \la 3 r_s$) region where the dynamo
dominates; we thus model the magnetic field profile crudely as a
series of three power laws.

As suggested by Falcke \& Markoff (2000), the soft X-ray spectrum is
probably due to the up-scattering of sub-mm and infrared photons by
synchrotron-self-Compton (SSC).  For a relativistic thermal
distribution of particles, the frequency of the peak of the SSC
spectrum will be $\sim 50 (k T / m_e c^2)^2 \nu_s$ where $T$ and
$\nu_s$ are the characteristic temperature and frequency,
respectively, of the magnetic bremsstrahlung emission.  Fitting this
to the observed spectrum suggests $T \sim 10^{11}$ K, consistent with
previous models (Coker \& Melia 2000) which did not, however,
calculate the SSC spectrum explicitly.

\section{Results of SSC Model}

\begin{figure}
\psfig{figure=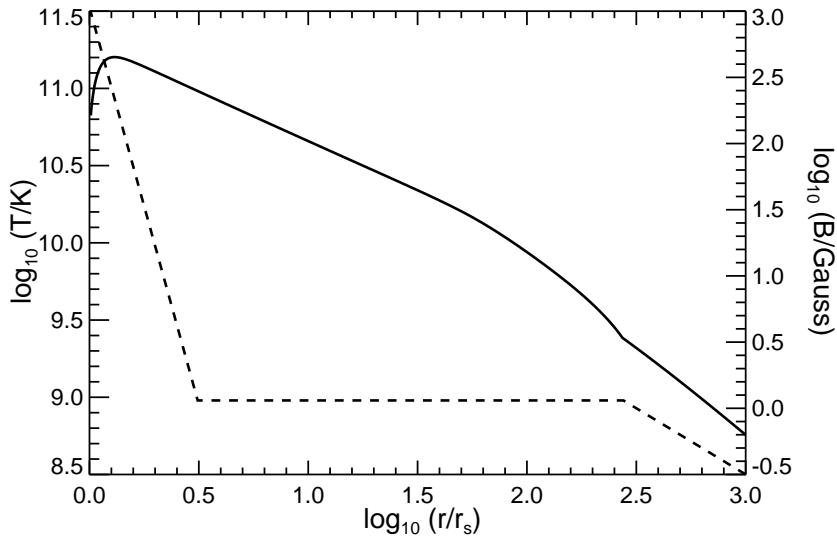,width=4.1in}
\caption{A plot of the
  temperature (solid line) and magnetic field (dashed line) versus
  radius for the best-fit model whose spectrum is shown in Figure 2.
  The central number density is $n_{\rm H} \simeq 6\times10^8$
  cm$^{-3}$ and corresponds to a mass accretion rate of
  $2\times10^{20}$ g s$^{-1}$.  For Sgr A*, $r_s = 8\times10^{11}$ cm.
  }
\end{figure}

\begin{figure}
  \psfig{figure=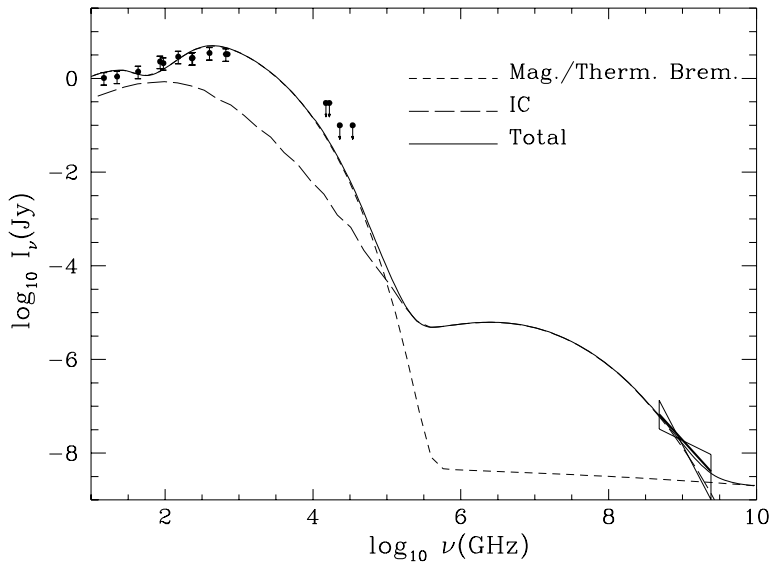,width=4.15in,angle=-90}
\caption{Best-fit combined thermal and magnetic bremsstrahlung, and
Compton up-scattered, spectrum for Sgr A*. Data references in Falcke
\& Markoff (2000).  The error bars on the radio points correspond to
10\% and are less than observed fluctuations (Zhao, Bower, \& Goss
2000).  The detected Chandra power-law includes estimated errors in
the spectral index (Baganoff et al. 2000). }
\end{figure}

Following discussions during the conference, we have developed a
spherical accretion model which, as a result of SSC, fits the flux and
the spectral shape of the X-ray observations while remaining
consistent with the radio data.  The input parameters are the magnetic
field profile and the initial conditions at the outer edge of the
flow.  The resulting radio and infrared emission spectra are then
isotropically up-scattered via SSC by integrating over spherical
shells, taking into account the self-absorption.

Our best-fit temperature and magnetic field profiles are presented in
Figure~1.  The best-fit mass accretion rate is
$3\times10^{-6}\;M_{\sun}$ yr$^{-1}$, a few times less than the upper
limit set by Quataert et al. (1999), while the peak temperature and
magnetic field are $3\times10^{11}$ K and $1000$ G, respectively.
Figure~2 shows the best-fit spectral components.  The model satisfies
the infrared upper limits, reproduces the X-ray spectral index and
flux, and is consistent with the variable radio data.

It is not clear why the required accretion rate is so much lower than
expected from HD simulations; perhaps there is an outflow, such as a
small jet, resulting in a radially dependent accretion rate.
Variability studies will be the key to discerning which model is
correct.  If there is no disk, variations in the flux from Sgr A* will
propagate from cm to X-ray frequencies in less than a day.  With a
proper accretion disk, the lag between cm and X-ray fluctuations will
be longer due to the reduced radial velocities.  For a jet-like
outflow, however, variations in the sub-mm will precede changes in the
cm and X-ray.


\begin{references}
Baganoff, F.K., et al. 2000, \apj, submitted.

Balbus, S. A. \& Hawley, J. F.  1991,
\apj, 376, 214

Bondi, H. \& Hoyle, F.
1944, \mnras, 104, 273 

Coker, R.F. \& Melia, F.  1997, \apjl, 488, L149

Coker, R.F. \& Melia, F.  2000, \apj, 534, 723

Falcke, H., et al. 1998, \apj, 499, 731

Falcke, H. \& Markoff, S.  2000,
A\&A, 362, 113

Genzel, R., et al. 
1996, \apj, 472, 153

Genzel, R., et al.  2000,
MNRAS, 317, 348

Ghez, A.M., et al. 2000,
Nature, 407, 349

Melia, F., Liu, S., \& Coker, R.F.  2000, \apj, submitted.

Najarro, F., et al. 1997,
\aap, 325, 700
%Najarro, F., Krabbe, 
%A., Genzel, R., Lutz, D., Kudritzki, R.P., \& Hillier, D.J.  1997, 
%\aap, 325, 700 

Quataert, 
E., Narayan, R., \& Reid, M.J.  1999, \apjl, 517, L101 

Predehl, P. \& 
Zinnecker, H. 1996, in ASP Conf. Ser. Vol 102, The Galactic Center,
ed. R. Gredel (San Francisco: ASP), 415

Reid, M.J., et al.
1999, \apj, 524, 816
%Reid, M.J., Readhead, A.C.S., Vermeulen, R.C., \& Treuhaft, R.N. 
%1999, \apj, 524, 816 

Serabyn, E., et al.
1997, \apjl,
490, L77
%Serabyn, E., Carlstrom, 
%J., Lay, O., Lis, D.C., Hunter, T.R., \& Lacy, J.H.  1997, \apjl, 
%490, L77 

Backer, D.C. \& 
Sramek, R.A. 1999, \apj, 524, 805 

Zhao, J.-H., Bower, GGG., \& Goss, W.M.  2000, \apjl, in press

\end{references}
\end{document}